\documentclass[times,10pt,twocolumn]{article}
\usepackage[latin1]{inputenc}
\usepackage{proof,amsmath,amsfonts,latexsym,amssymb,color}
\usepackage{enumerate}

\def\parag#1{{\smallskip \bf #1.}}
\def\liste#1{\mathcal{L}(#1)}
\def\fm{\multimap}
\def\xreduce#1{\xrightarrow{#1}}
\newcommand{\la}{\lambda}
\newcommand{\al}{\alpha}
\newcommand{\lambdak}{\lambda_{LA \, \mathbb{K}}}
\def\mes#1{|#1|}
\def\size#1{size{(#1)}}
\newcommand{\LAL}{\mbox{\bf LAL}}
\newcommand{\NN}{\mathbb{N}}
\newcommand{\ZZ}{\mathbb{Z}}
\newcommand{\KK}{\mathbb{K}}
\newcommand{\RR}{\mathbb{R}}
\newcommand{\CC}{\mathbb{C}}
\newcommand{\mylet}[3]{\mbox{$\mathtt{let} \,  #1 \,\mathtt{be}\, #2 \,\mathtt{in}\, #3$}}
\def\pgf{\mbox{\S}}
\newcommand\remove[1]{}

\newtheorem{lemma}{Lemma}
\newtheorem{theorem}{Theorem}
\newtheorem{definition}{Definition}

\begin{document}
\title{An Embedding of the BSS Model of Computation in\\
 Light Affine Lambda-Calculus
\thanks{Work partially supported
by Projects ``Interaction and Complexity'' (cooperation project 2004-2006: CNR, Italy - CNRS, France), GEOCAL (ACI), NO-COST (ANR).}}

\remove{
\author[P.~Baillot]{Patrick Baillot}
\address{
Laboratoire d'Informatique de Paris-Nord,
Institut Galilee, Universite Paris XIII
99 av. Jean-Baptiste Clement, 93430 Villetaneuse,
 France}
\author[M.~Pedicini]{Marco Pedicini}
\address{
Istituto per le Applicazioni del Calcolo ``M.~Picone'', CNR, 
Viale del Policlinico 137, 00161 Rome, Italy}
}

\author{Patrick Baillot\\
LIPN, CNRS / Universit\'e Paris 13, France\\
patrick.baillot@lipn.univ-paris13.fr
\and
Marco Pedicini\\
IAC, Consiglio Nazionale delle Ricerche, Roma, Italy\\
marco@iac.rm.cnr.it
}
\date{} 

%
\maketitle

\begin{abstract}
  This paper brings together two lines of research: implicit
  characterization of complexity classes by Linear Logic (LL) on the
  one hand, and computation over an arbitrary ring in the
  Blum-Shub-Smale (BSS) model on the other. Given a fixed ring
  structure K we define an extension of Terui's light affine
  lambda-calculus typed in LAL (Light Affine Logic) with a basic type
  for K. We show that this calculus captures the polynomial time function
  class FP(K): every typed term can be evaluated in polynomial time
  and conversely every polynomial time BSS machine over K can be
  simulated in this calculus.
\end{abstract}


\section{Introduction}

\parag{BSS computation} The Blum-Shub-Smale (BSS) model was introduced
as an extension of the classical model of Turing machines to describe
computations on an arbitrary ring or field $K$ (\cite{BSS89}; see also
\cite{BCSS98}).  The idea is basically to consider an idealized
machine which can store elements of the ring and perform on them a
certain number of operations or tests at unary cost.  The initial
interest was on computation over reals and one motivation was to get a
framework to reason about complexity of algorithms from numerical
analysis and applied mathematics. Complexity classes analogous to the
ones of the classical setting have been defined and this setting
subsumes the classical one in the case $K$ is taken to be $\ZZ \slash
2 \ZZ$ (with expected boolean operations).  Moreover this approach was
later extended to arbitrary logical structures (\cite{Poizat95}).

 One might object that the BSS model is questionable from the point
of view of physical realization: performing equality tests on real numbers
at unary cost for instance is problematic. Anyway observe  that it provides
a setting which is compatible with the common way of handling complexity
in numerical analysis an symbolic computation; thus we think it is a 
relevant model.

\parag{Implicit computational complexity} In the classical
computational pa\-ra\-digm some work has been done to characterize
functions of various complexity classes without reference to a machine
model and explicit resource bounds. This line of research, Implicit
Computational Complexity (ICC), has been developed using various
approaches such as recursion theory (\cite{BellantoniCook,Leivant94}),
lambda-calculus (\cite{LeivantMarion93}) or logic
(\cite{Girard98}). On the practical side it has yielded techniques for
automatically or partially automatically inferring complexity bounds on
programs (\cite{Hofmann99,Jones01,MarionMoyen00}).

Linear logic (LL, \cite{Girard87a}) has provided one line of research
in ICC which fits in the proofs-as-programs paradigm: variants of LL
with strict resource duplication disciplines such as Light Linear
Logic (\cite{Girard98}, or its variant Light Affine Logic,
\cite{Asperti98}) or Soft Linear Logic (\cite{Lafont02}) capture
deterministic polytime computation. Light Affine Logic (LAL) has in
particular been studied using specific term calculi
(\cite{Asperti98,Roversi00}); among these, Terui's \textit{light affine
  lambda-calculus} enjoys good properties and has
allowed to prove new properties on LAL (like the strong polytime bound, see
\cite{Terui01}). Some
advantages of the light logics approach are the fact that it allows
higher-order computation (also the case in 
\cite{Hofmann00,BellantoniNigglSchwichtenberg00}), polymorphism and enables to
define new datatypes (as in system F). It fits also well with program
extraction from termination proofs: in \cite{Girard98,Terui04}) a
naive set theory is presented in which the provably total functions
are exactly the polytime functions; the witness programs are extracted
from proofs as light affine lambda-calculus terms.

\parag{ICC and BSS} An extension of ICC to the BSS model was proposed
by Bournez \textit{et al.} in \cite{Marion-BSS}: this article
characterizes in particular by means of \textit{safe recursion} the
class FP(K) over an arbitrary structure. Recall that safe recursion
was introduced by Bellantoni and Cook (\cite{BellantoniCook}) as a
restriction of primitive recursion based on the distinction between
two classes of arguments (normal and safe) and characterizing the
(classical) class FP.  In the BSS case the recursion considered is on
the structure of lists over $K$. 
 Their approach was extended to
other complexity classes such as PAR and the polynomial hierarchy in
\cite{BCJM03} thus demonstrating the relevance of ICC tools to the BSS
framework.

\parag{Our goal and contribution} In the present work we use light
affine logic and light affine lambda-calculus to provide a new
characterization of the class $FP(K)$ of deterministic polynomial time
functions over $K$. In the  long term we wish to develop a theoretical
language to write feasible algorithms on an arbitrary ring and to
allow formal reasoning on these algorithms.  We think that
lambda-calculus and LAL offer two main advantages in this perspective:
\begin{itemize}
\item higher-order: in numerical analysis algorithms functions have a
  first-class status, and one naturally handles higher-order
  functionals; thus it is an important point to have a language which
  includes higher-order;
\item proofs: computation over $\RR$, $\CC$ or other rings or fields is
a framework in which we would certainly like to be able to manage together
mathematical proofs and programs in an integrated way;
Light linear logic is interesting in this respect because it provides a setting
which can accommodate program extraction from proofs. 
\end{itemize}
Finally, some semantic interpretations of Light linear logic have 
been given, both for semantics of formulas (phase spaces,
\cite{KOS03}) an for semantics of proofs (in games, \cite{MO00} or
coherent spaces \cite{Baillot04}). Thus the present work provides a first
step from
which semantic approaches for the study of BSS polytime functions  can be 
considered.

Concretely, our extension of light lambda-calculus is very simple: to
the type language (LAL) we just add a basic type for $K$ and to the
term language some constants for the elements of $K$ and for the
operations and relations of $K$ (a bit as in the language PCF with the
type of integers for instance). The contribution of the present paper
is then to show the validity of this approach:
\begin{itemize}
\item we show that any term on lists over $K$ in this language denotes
  an $FP(K)$ function;
\item we show that BSS polytime machines can be simulated, hence all $FP(K)$
functions can be programmed. 
\end{itemize}
\parag{Outline of the paper} In section \ref{LAL} we recall Light affine logic and
light affine lambda-calculus and in section \ref{BSSMachines} the BSS
model. In section \ref{lightK}, we define our extension $\lambdak$ of light
affine lambda-calculus to a structure $K$; then we show that the terms
can be reduced in polynomial time (section \ref{boundlightK}) and
conversely that all ptime BSS machines over $K$ can be simulated in
$\lambdak$ (section \ref{encodingBSS}).

\section{\LAL\  and  $\lambda_{\LAL\ }$}\label{LAL}

The formulas of Intuitionistic Light affine logic (LAL) are given by
the following grammar:
$$A, B := \al \; | \; A \fm B \;|\; !A \; |\; \pgf A \;|\; \forall \al. A$$
  
The modalities $!$, $\pgf$, called \textit{exponentials} are used to
control duplication.  An erasure map $(.)^-$ from LAL formulas to
system F types is given by:

  $(A \fm B)^-= A \rightarrow B$, $(!A)^-= (\pgf
 A)^-= A^-$, $(\forall \al.A)^-=\forall \al.A^-$.
 
 Following Terui (\cite{Terui01}), we consider
$\lambda_{\mbox{\sc LAL}}$ a typed lambda-calculus with types of
intuitionistic light affine linear logic. This calculus has explicit
constructs for handling $!$ and $\pgf$. Its terms are defined by the grammar:
  $$t, u ::= x | \lambda x\, t
| (t)u | !t | \mylet{u}{!x}{t} | \pgf t | \mylet{u}{\pgf x}{t} $$
In typing judgments, besides ordinary LAL formulas we will use
!-\textit{discharged} and $\pgf$-\textit{discharged} formulas, of the
form $[A]_{\dagger}$ with respectively $\dagger=!$ or $\pgf$.
Discharged formulas have a temporary status, they cannot be applied
any connective and are only a technical artifact to manage structural
 rules (contractions) in a convenient way.
 
 The typing rules are now given on Figure \ref{LALtypingrules}.

\begin{figure*}[ht]
$$
\begin{array}{cc}
\infer[Id]{x:A \vdash x:A}{} & 
\infer[Cut]{\Gamma_1, \Gamma_2 \vdash t[u/x]:C}{\Gamma_1 \vdash u:A \qquad x:A, \Gamma_2 \vdash t:C}\\
\\
\infer[Weak]{\Delta, \Gamma\vdash t: C}{\Gamma \vdash t: C} & 
\infer[Cntr]{z:[A]_! , \Gamma \vdash t[z/x, z/y]:C}{x:[A]_!, y:[A]_!, \Gamma \vdash t:C}
\end{array}
$$
$$
\begin{array}{cc} 
\infer[\multimap_l]{\Gamma_1 ,y :A_1 \multimap A_2 , \Gamma_2 \vdash t[(y)u /x] : C
}{\Gamma_1 \vdash u : A_1 & x: A_2 , \Gamma_2 \vdash t:C} & 
\infer[\multimap_r]{\Gamma \vdash \lambda x\, t: A_1 \multimap A_2}{x:A_1 , \Gamma \vdash t: A_2}
\end{array}
$$
$$
\begin{array}{cc} 
\infer[\forall_l]{x:\forall \alpha A,\Gamma \vdash t:C}
{x:A[B/\alpha],\Gamma \vdash t:C} & 
\infer[\forall_r]{\Gamma \vdash t:\forall \alpha A}
{\Gamma \vdash t : A} \\
& (\alpha \mbox{ not free in $\Gamma$})\\
\\
\infer[!_l]{y : !A , \Gamma \vdash \mylet{y}{!x}{t}:C}
{x:[A]_!,\Gamma \vdash t:C} & 
\infer[!_r]{x: [B]_! \vdash !t : !A}{x:B \vdash t:A} 
\\
&\mbox{ with a possibly empty context.}\\
\\ 
\infer[\pgf_l]{y : \pgf  A , \Gamma \vdash \mylet{y}{\pgf  x}{t}:C}{x: [A]_{\pgf}  , \Gamma \vdash t:C } & 
\infer[\pgf _r]{ [\Gamma]_! , [\Delta]_{\pgf} \vdash  \pgf t : \pgf  A}{\Gamma , \Delta \vdash t:A} \\
 & \mbox{ with $\Gamma$ and $\Delta$ possibly empty.}
\end{array}
$$
\caption{LAL typing rules.}\label{LALtypingrules}

\end{figure*}

Note that the typing rules are here given in a sequent calculus style
(with right and left introduction rules); a natural deduction
presentation could also  have been used.

 The most important rule to notice is
$!_r$: as $Cntr$ is performed only on $!$-discharged variables, during
reduction only $!$ typed terms will be duplicated; the $!_r$ rule
ensures that duplicable terms have \textit{at most one occurrence} of free
variable. This is one of the keys that ensure the polynomial bound
for  the reduction of these terms (\cite{Terui01}). Another
important point is a stratification property ensured by the $!$ and $\pgf$
connectives: in particular note that to $!$ discharge a variable $x$
(thus making it contractible) one has to apply a $!_r$ or a $\pgf_r$ rules
and add an exponential to the type of the term.

Actually, in \cite{Terui01} terms are defined as a subclass of
\textit{pseudo-terms} satisfying some syntactical conditions. We could
do the same here but as we will only consider typed terms this is not
necessary (all well-typed pseudo-terms are terms).

The reduction relation is defined as the contextual, reflexive and transitive
closure of the relation given on Figure \ref{reductionsteps}.
\begin{figure*}[ht]
\fbox{
$
\begin{array}{c}
(\la x. t)u \xreduce{(\beta )}  t[u/x]\\
 \mylet{!u}{!x}{t} \xreduce{(!)} t[u/x]  \\
 \mylet{\pgf u}{\pgf x}{t} \xreduce{(\pgf)} t[u/x] \\
 \mylet{(\mylet{u_1}{\dagger_1 x}{t_1})}{\dagger_2 y}{t_2} \xreduce{(com1)} 
\mylet{u_1}{\dagger_1 x}{(\mylet {t_1}{\dagger_2 y}{t_2})} \\
 ( \mylet{u_1}{\dagger_1 x}{t_1})t_2 \xreduce{(com2)} 
 \mylet{u_1}{\dagger_1 x}{(t_1)t_2} \mbox { where } \dagger_i= ! \mbox{ or }\pgf, \mbox{ for } i=1,2
\end{array}
$
}
\caption{Reduction rules}\label{reductionsteps}
\end{figure*}
Actually the $\beta$ rule is a linear beta reduction step and only the $!$ rule can 
cause duplications.
 
Terms of $\lambda_{\mbox{\sc LAL}}$ should in fact be seen as ordinary lambda-terms
with extra information on sharing and stratification given by the 
$!$ and $\pgf$ constructs. By erasing this information from a term $t$
 we get an ordinary
 lambda-term $t^-$ which denotes the same function as $t$:
$$\begin{array}{lcl}
 (!t)^- &=& (\pgf t)^-=  t^-,\\
x^-&=& x, (\lambda x\, t)^-=\lambda x\, t^-, [(t)u]^-= (t)^- u^-,\\
\end{array}
$$
$$(\mylet{u}{\dagger x}{t})^-= t^-[u^-/x].$$
If $\Gamma \vdash_{LAL} t:A$ we then have in system F: $(\Gamma)^-
\vdash_{F} t:A^-$.  Moreover if $t$ is a $\lambda_{\mbox{\sc LAL}}$
term and $t \xreduce{} t'$, then we have with
ordinary beta reduction: $t^- \xreduce{\star} t'^-$
  
 We could in fact instead of $\lambda_{\mbox{\sc LAL}}$ have used 
ordinary lambda-terms typed in DLAL (see \cite{BaillotTerui04}),
 a system which is essentially a fragment of LAL. The properties in the rest
of this paper could have been proved in the same way. 

\subsection{Syntactic sugar: lambda calculus macro definitions}

\subsubsection{Tensor}

We consider $\otimes$ as a defined construct. On types we set:
$$A \otimes B= \forall \al (A \fm B \fm \al) \fm \al.$$
On terms we define:

$
\mylet{u}{x \otimes y}{t}= (u) \la x \la y. t$

$t_1 \otimes t_2 = \la y (y)t_1 \; t_2.$

Then we have the following typing rules:
$$
\begin{array}{c} 
\infer[\otimes_l]{z: A\otimes B, \Gamma \vdash \mylet{z}{x\otimes y}{t}:C}
{x : A , y : B,\Gamma \vdash t:C}\\
\infer[\otimes_r]{ \Gamma , \Delta \vdash   t\otimes u : A\otimes B}{\Gamma \vdash t:A &  \Delta \vdash u:B} .\\
\end{array}
$$
and the reduction rule:
$$\mylet{(u_1 \otimes u_2)}{(x_1 \otimes x_2)}{t} \xreduce{(\otimes)} t[u_1/x_1, u_2/x_2].$$

In the sequel we will use as a short-hand compound patterns 
such as for instance $x_1 \otimes x_2 \otimes x_3$ or $( \pgf x_1 ) \otimes x_2$,
for which the $\mathtt{let}$ constructs are definable from the $\mathtt{let}$ constructs
for $!$, $\pgf$, $\otimes$.
 
Let us also denote by $\lambda x\otimes y. \,t$ 
the term $\lambda z\, \mylet{z}{x\otimes y}{t}$. Hence we have:
$$(\lambda x\otimes y .\, t)u\otimes v \xreduce{} t[u/x, v/y].$$

\subsubsection{Integers and Booleans encodings}

Tally integers are given by the type:
$N =\forall \al. !(\al \fm \al) \fm \pgf (\al \fm \al)$.

Booleans are defined by the type $Bool = \forall
\alpha . (\alpha \multimap \alpha \multimap \alpha) $:

$\mathtt{true} =
\lambda x \lambda y. y$ and $\mathtt{false} = \lambda x \lambda y. y$. 


We define a  term for conditional:

$\mathtt{if \mbox{ $b$ } then \mbox{ $u_1$ } else \mbox{ $u_2$}} =  
(((b) \lambda x_1 \dots \lambda x_n . u_1) \lambda x_1 \dots \lambda x_n . u_2 )x_1 \dots x_n$

with  the typing rule:
$$
\infer[ite]{
\Gamma \vdash \mathtt{if \mbox{ $b$ } then \mbox{ $u_1$ } else \mbox{ $u_2$}} : A
}{
\vdash b : Bool & \Gamma \vdash u_1 : A  & \Gamma \vdash u_2 : A
}
$$

\section{BSS Machines over $\KK$}\label{BSSMachines}

Recall that a (classical) Turing machine over a finite alphabet $A$ and
finite set of states $Q$ is given by a function
$$\mu : Q \times \tilde A \to Q' \times \tilde A \times \{ -1 , 0, 1 \}$$
 where $\tilde A := A \cup \Box$ and $Q' := Q \cup \{q_F^0 , q_F^1\}$ with
 $q_F^0 , q_F^1 \not\in Q$ are respectively called 
rejecting and accepting state.

Let $(K, +, *, 0,1)$ be a ring. A \textit{structure}
on $K$ is a tuple:
$$\KK = ( K , op^{k_1}_1, \dots ,
op^{k_n}_n, \rho^{s_1}_1, \dots \rho^{s_m}_m),$$
 where each
$op^{k_i}_i$ is a polynomial function (operation) over $K$ of arity
$k_i$ and each $\rho_i$ is a predicate over $K$ of arity $s_i$. We
assume one of the predicates, say $\rho_1$, is the equality. In the case
where $K$ is a field the $op^{k_i}_i$ can be defined by rational
functions instead of polynomials. Operators of arity $0$ are constants.

Two examples of structures are:
$$\begin{array}{ccl}
 {\KK}_1 &=& ( \RR , +, -, *, (c_i)_{i \in \RR }, =, \leq ), \\
 {\KK}_2 &=& ( \{0,1 \} , \vee, \wedge, 0,1, =).
\end{array}
$$

 A \textit{BSS machine} over $\KK$ (\cite{BCSS98}) is a generalization of
Turing machines that we shall describe below.

For a given structure  $\KK$ we denote by $K_\infty = K^\ZZ$
and by $$K^\infty = \bigcup_{i=1}^\infty K^m\quad \quad
\mbox{where $\displaystyle{
K^m =\{ (x_1, \dots , x_m) |  x_i \in K\}}$.}$$

A machine has a finite set of states $Q$ and for each state $q \in Q$
only one of the following kinds of actions can be performed:
\begin{itemize}
\item (computation) at this step we aim to compute the value of 
one of the operations $op_i$ using the first $k_i$ elements 
of $K^\infty$, the result is then stored in place of the 
current position.
\item (branch)
at this step we aim to compute the value of 
one of the relations $\rho_i$ using the first $s_i$ elements 
of $K^\infty$, the result is used to select a state.
\item (shift) 
this last type of action a BSS machine can perform 
corresponds to the movement of the head of the machine
(on to the left or on to the right).
\end{itemize}

\begin{definition} 
Given a structure $\mathbb{K}$, a machine over $\mathbb{K}$ is a function
$\mu : Q \to  \mathcal{F}$,
where 
$$\mathcal{F}=\bigcup_{i\in \NN} (\tilde K^{i} \to Q' \times \tilde K \times \{ -1 , 0, 1 \}).$$
where $Q'=Q\cup \{q_{F}^1,q_{F}^0\}$ and $q_{F}^1,q_{F}^0\not\in Q$.

Every state $q$ in $Q'$ (also called node) can be of one of the
following five types:
\emph{computation}, \emph{branch} or \emph{shift} as described above, 
\emph{input} or \emph{output}.

For every node in $Q$ the corresponding action is 
determined by $q$, it depends on $n_q$ elements of $K$
and gives as a result a triple $Q' \times \tilde K
\times \{ -1 ,0, 1 \}$:
 $$\mu(q) : \tilde K^{n_q} \to Q' \times \tilde K
\times \{ -1 ,0, 1 \}$$
 
\begin{itemize}
\item (computation) 
$$\mu(q)(k_1, \dots, k_{n_q})=( q', op_i( k_1, \dots, k_{n_q} ), 0)$$
where $op_i$ is determined by $q$;
\item (branch) 
$$\mu(q)(k_1, \dots, k_{n_q})=(q_b , k_1, 0)$$ 
where $b = \rho_i(k_1, \dots, k_{n_q})$
and the relation $\rho_i$ to be applied is determined by $q$;
\item (shift)
$$\mu(q)(k_1, \dots, k_{n_q})=( q' , k_1 , m_q)$$
and $q'$ only depends on $q$ (note that here dependency on
$k_1, \dots, k_{n_q}$ is formal).
\end{itemize}

We have exactly one input node denoted by $q_0 \in Q$, 
and two distinguished output nodes denoted by $q_{F}^1 \in Q'$
and $q_{F}^0 \in Q'$. 
\end{definition}

All the actions modify the current configuration of the 
machine: let us define a configuration of the machine at a given time
as a triple $(f,p,q)$ where $f \in \KK^\ZZ$, $p\in \ZZ$ and $q\in Q'$.

In the encoding part since we cannot represent by a term the infinitary $f$ 
 but only a finite part of it, we will represent such a
configuration as a triple $\langle f^-, f^+, q \rangle$, where $f^- =
(f(-1), f(-2), \dots ,f(-n^-))$ if $ f(-k) = \Box$ for all $k\ge n
^-$, and analogously $f^+ = (f(0), f(1), f(2), \dots ,f(n^+) )$ if $
f(k) = \Box$ for all $k\ge n ^+$.

The initial configuration associated to $w \in K^\infty$ is
$(f_w,1,q_0)$ where $f_w \in K_\infty$ is
$$f_w(i) = \begin{cases}
w_i &\mbox{for all $1\le i \le |w|$}  \\
 \Box & \mbox{otherwise}.
\end{cases}
$$
 
A transition from the configuration $(f,p,q)$ gives the
configuration $(f',p+m,q')$ if $$\mu(q)(f(1), \dots,
f(n_q))=(q',k,m)$$ and $f'(i)=f(i)$ for every $i\neq p$ and
$f'(p)=k$.
 
Since $\mu$ is not defined on output nodes, they correspond to the
end of the computation.

An input word $w$ is \emph{accepted} by $\mu$ if the machine starting
from the input node evolves according to the transitions specified by its
nodes and eventually reaches the $q_{F}^1$ output node.

A language $L\subset K^\infty$ is said to be \emph{recognized} by
$\mu$ if and only if it corresponds to the words which are accepted by $\mu$.

A machine $\mu$ computes a function $g : K^\infty \to K^\infty$ if for
any $w\in K^\infty$, $\mu$ evolves from the initial configuration $(f_w,
1, q_0)$ to a terminal configuration $(f_{g(w)}, 1, q_F^1)$ if $f$ is
defined on $w$ and the computation does not end if $f$ is not defined
for $w$.


\section{A light lambda-calculus for the structure $\KK$}\label{lightK}

In order to manage arities we consider a variant 
of the definition of BSS machines; for a given structure 
$\KK$, let us consider the maximal arity $p$ for $op_i$ and 
$\rho_i$ in $\KK$:
 $$p = \max \displaystyle{(\max_{1\le i \le n}{k_i} , \max_{1\le i \le m} s_i)}.$$

Then we consider an algebraic structure $\KK^{up}$ with 
all the operations and relations of arity $p$ defined as follows:
$$op_i'(x_1 , \dots , x_p) = op_i(x_1 , \dots , x_{k_i})$$
and
$$\rho_i'(x_1 , \dots , x_p) = \rho_i(x_1 , \dots , x_{s_i}).$$

For the algebra $\KK^{up}$ we have a transition
step with  uniform number of elements of the tape:
$\mu(q) : K^{p} \to Q' \times \tilde K \times \{ -1 , 0, 1 \}$.
Below we suppose $\KK$ of uniform arity $p$.

We extend types of LAL with a basic type $\kappa$ for elements of
$K$. We add to the language constants $\star$ (used as empty list
symbol), $\underline{k}$ for each $k \in K$ and $\mathtt{dup}$ (for
duplication), $\mathtt{op}_i$ ($1 \leq i \leq n$), $\mathtt{\rho}_i$
($1 \leq i \leq m$) for each operation and relation of the structure.
The new term language $\lambdak$ is thus given by:
\begin{align*}
t, u ::=& x  |\lambda x\, t | (t)u |\\
&  !t | \mylet{u}{!x}{t} | \pgf t | \mylet{u}{\pgf x}{t} | \\
& t_1 \otimes t_2 | \mylet{u}{x\otimes y}{t} | \mathtt{dup} |\\
& \underline{k} | \star | \mathtt{op}_i  | \mathtt{rho}_i,
\end{align*} 
for every $k \in K$. The new typing rules for the constants are given
on Figure \ref{typingconstants} where $\kappa^p$ denotes
$\underbrace{\kappa \otimes \dots \otimes \kappa}_{p}$.

\begin{figure*}[ht]
\begin{center}\fbox{
$
\begin{array}{ccc}
\infer[\kappa]{\vdash \underline{k} : \kappa }{}&
\infer[\star]{\vdash \star : \kappa }{}&
\infer[dup]{\vdash \mathtt{dup} : \kappa  \multimap \kappa \otimes \kappa}{}\\
&&\\
\infer[op_i]{\vdash \mathtt{op}_i : \kappa^p \multimap \kappa}{}&
\infer[\rho_i]{\vdash \mathtt{rho}_i :\kappa^p \multimap Bool}{}&\\
\end{array}
$
}\end{center}
\caption{Typing rules for constants.}\label{typingconstants}
\end{figure*}

For these constants we consider associated reduction rules given on Figure~\ref{newreductionrules}.
We will denote by $\xreduce{}$ the resulting new reduction relation and by 
$\xreduce{\star}$ its reflexive and transitive closure.
\begin{figure*}[ht]
\begin{center}
\fbox{
$
\begin{array}{ll}
(\mathtt{dup})\underline{k} \xreduce{(dup)}  \underline{k} \otimes \underline{k}\\
(\mathtt{op}_i) \underline{k_1} \dots \underline{k_p} \xreduce{(op)}
\underline{k} & \mbox{ if } op_i(k_1, \dots , k_{k_i})=k,\\
(\mathtt{rho}_i) \underline{k_1} \dots \underline{k_p} \xreduce{(rho)} b  &\mbox{ where 
if }\rho_i(k_1, \dots , k_{s_i}) \mbox{ holds (resp. does not hold) then }\\
  &  b=\mathtt{true}
\mbox{ (resp. $b=\mathtt{false}$)}
\end{array}
$
}
\end{center}
\caption{Reduction rules for constants.}\label{newreductionrules}
\end{figure*}
Note the reduction $(dup)$ is performed only when the argument is a value.
Given a type $A$ we will consider the usual type for lists of elements
of $A$: $\liste{A}= \forall \alpha. !(A \fm \al \fm \al) \fm \pgf (\al
\fm \al)$. So for lists over $K$ we have the type $\liste{\kappa}$. We
denote by $nil$ the empty list. Recall that $\liste{A}$ allows defining
for any $B$ a $\mathtt{fold_B}$ map with type:
 $$\mathtt{fold_B}: !(A \fm B \fm B) \fm \pgf B \fm \liste{A} \fm \pgf B.$$

As in previous work on light logics (see \cite{Girard98} and
\cite{AspertiRoversi2002}) we will represent functions on lists by
terms of type $\liste{A} \fm \pgf^n \liste{A}$, where $n$ is an
integer.

\section{The calculus $\lambdak$ is polytime}\label{boundlightK}

We want to show that terms of $\lambdak$ can be reduced in a
polynomial number of steps. For that we adapt Terui's proof of weak
polystep normalization for light affine lambda-calculus in
\cite{Terui01}, which in turn follows \cite{Girard98}.

We consider a measure for terms of  $\lambdak$ defined by: 
 $$\begin{array}{ll}
\mes{x}=1, & \\
\mes{\underline{k}}= 1, & \mes{\la x. t}=\mes{t}+1,\\
\mes{op_i}= 2, & \mes{(t)u}=\mes{t}+\mes{u}+1,\\
\mes{\mathtt{rho}_i}=4, &  \mes{ \dagger t}=\mes{t}+1, \mbox{ for }\dagger=!, \pgf\\
\mes{\mathtt{dup}}= 5, & \mes{\mylet{u}{\dagger x}{t}}= \mes{t}+\mes{u}+1,\\
& \qquad \qquad  \mbox{ for }\dagger=!, \pgf
\end{array}
$$

We denote by $\size{t}$ the number of nodes of the syntactic tree of
$t$; therefore $\size{t} \leq \mes{t}$. We have:
\begin{lemma}
  If $t \xreduce{(r)} t'$ and $(r) \neq (!), (com \, i)$ then
  $\mes{t'} < \mes{t}$. If $(r)=(com \, i)$ then $\mes{t'}=\mes{t}$.
\end{lemma}

Let $t \xreduce{\sigma \; \star} t'$ be a reduction sequence of $t'$.
The \textit{length} $\mes{\sigma}$ of the reduction $\sigma$ is its
number of steps.

We say a reduction of a term $t$ is \textit{standard} if it is
obtained in the following way:
$$t=t_0 \xreduce{\star} t_1 \xreduce{\star} \dots t_i \xreduce{\star}
t_{i+1} \dots \xreduce{\star} t_n$$
where:
\begin{itemize}
\item sequences $t_{2j} \xreduce{\star} t_{2j+1}$ consist only of non $(!)$ reduction steps at depth $j$,
 \item sequences $t_{2j+1} \xreduce{\star} t_{2j+2}$ consist only of  $(!)$ reduction steps at depth $j$. 
\end{itemize}  

With the notion of measure chosen the proofs of the other lemmas of \cite{Terui01}
remain valid for $\lambdak$ and we get in the same way:
\begin{theorem}
  Let $t$ be a $\lambdak$ term of depth $d$ and $\sigma$ be a standard
  reduction $t \xreduce{\sigma \; \star} u$; then $\mes{u} \leq
  \mes{t}^2$ and $\mes{\sigma} \leq \mes{t}^{2^{d+1}}$.
\end{theorem} 
Moreover, each reduction step on a term $v$ can be simulated on a
BSS machine over $\KK$ in a number of steps proportional to
$\size{v}^2$, so to $\mes{v}^2$. Hence each step of the standard
reduction $\sigma$ can be simulated in time $O((\mes{t}^{2^d})^2)$, so
$O(\mes{t}^{2^{d+1}})$. Therefore $\sigma$ can be simulated in time 
$O(\mes{t}^{2^{d+1}}.\mes{t}^{2^{d+1}})=O(\mes{t}^{2^{d+2}})$, so 
polynomial in $\mes{t}$ (at fixed depth $d$).

It follows then that:
\begin{theorem}
  If $f$ is a function on lists over $\KK$ representable by a
  $\lambdak$ term, then $f$ belongs to the class $FP (\KK )$.
\end{theorem}

\section{Encoding of ptime BSS machines}\label{encodingBSS}

The first difference with Roversi's encoding of classical Turing
machines (\cite{AspertiRoversi2002}) is due to the fact that in
the BSS-model the choice between branching, computation or shift depends
exclusively upon the current state; this simplifies the construction
of the transition function which usually is given as a bidimensional
table, whereas here we encode it as an array whose elements are
selected by the current state.

\subsection{States}

If $Q$ is the set of states of the machine $\mu$:
$$Q = \{q_0 , \dots , q_d\}$$
we encode the state $q_i$ by the term $\mathtt{q}_i = \lambda x_0 \lambda x_1 \dots \lambda x_d \lambda v (x_i) v$. In fact, the 
term $\mathtt{q}_i$ is a selector to extract from a table
a function of type 
$$\alpha \otimes \kappa^p \otimes \kappa^p \otimes \alpha \multimap
\alpha \otimes\alpha \otimes Q.$$
This function will be obtained by
composing either the terms representing $op_i$ or $\rho_i$ with a term which transforms the
current state in the next one.

So we have that the type of states is
$$Q = \forall \alpha \forall \beta (\underbrace{(\alpha \multimap \beta ) \otimes \dots \otimes (\alpha \multimap \beta)}_{d-times} \otimes \alpha ) \multimap \beta.$$

\subsection{Configurations and transitions}

As already introduced in section \ref{BSSMachines}, a configuration is
a triple given by the left (half)tape (or negative tape) by the right
tape (or positive tape) and by the current state.
\begin{multline*}
\langle f^-, f^+, q \rangle =
\lambda g \lambda x \lambda x' ((g)k_1^-)((g)k_2^-) \dots ((g)k_{n^-}^-)x  \otimes \\
 \otimes (((g)k_1^+)((g)k_2^+) \dots ((g)k_{n^+}^+)x' \otimes q_i$$
\end{multline*}
when
$f^- = (k_1^-, k_2^-, \dots , k_{n^-}^-)$,
$f^+ = (k_1^+, k_2^+, \dots , k_{n^+}^+)$
with $k_i^\epsilon : \kappa$ and $q :  Q$.
The type of configurations is then:
$$ C = \forall \alpha !(\kappa \multimap \alpha \multimap \alpha )
\multimap \pgf \alpha \multimap \pgf \alpha \multimap \pgf (
\alpha\otimes\alpha\otimes Q)
$$

 Note that to execute an action the machine might have to look ahead
on the tape the next $p$ squares: in the case of (computation) or
(branch) action this is needed to fetch the $p$ arguments on which to
apply $op_i$ or $\rho_i$. It is convenient for that to have a
\textit{window} of length $2p$ of elements directly accessible
(representing the $p$ elements on the right and on the left of the
head). Such a window will have type $\kappa^p \otimes \kappa^p$.

Thus for the encoding we proceed in two steps:
\begin{itemize}
 \item first, from a configuration we produce a \textit{configuration
with window}, of type: 

$ CW = \forall \alpha !(\kappa \multimap \alpha \multimap \alpha )
\multimap \pgf \alpha \multimap \pgf \alpha \multimap \pgf (
\alpha\otimes \kappa^p \otimes \kappa^p \otimes \alpha\otimes Q).$

\item second, on a configuration with window we perform a transition step
(with an action determined by the state) which produces a new configuration.
\end{itemize}
The first step will be done by a term $\mathtt{c2cw}:C \fm CW$; using this term
in the second step we will give a term $\mathtt{c2c}:C \fm C$.

 Let us first consider the term $c2cw:C \fm CW$: applied to a
configuration $\langle f^-, f^+, q \rangle$, $\mathtt{c2cw}$ will
yield a configuration with window:
 $$\langle h^-, (k_1^-, k_2^-, \dots , k_{p^-}^-), (k_1^+, k_2^+, \dots , k_{p^+}), h^+, q \rangle$$
where:
\begin{align*}
f^- &= (k_1^-, k_2^-, \dots , k_{n^-}^-),& \; f^+ &= (k_1^+, k_2^+, \dots , k_{n^+}^+),\\
h^- &= (k_2^-, \dots , k_{n^-}^-),& \; h^+ &= (k_2^+, \dots , k_{n^+}^+),
\end{align*}
if $n^+\geq p$ and $n^-\geq p$. In the case where $n^+< p$ or $n^-< p$
the missing values for the window are replaced by $\star$.

 The term $\mathtt{c2cw}$ is defined by an iteration:
\begin{multline} 
\mathtt{c2cw} = \lambda c \lambda g \lambda x \lambda x'.
\mathtt{let}
(c) \mathtt{step}\, \mathtt{base}^- \, \mathtt{base}^+ \\ 
\mathtt{be \; }
\pgf ( (b_1 \otimes \vec{k_1} \otimes l_1) \otimes (b_2 \otimes \vec{k_2} \otimes l_2))\otimes q\\
\mathtt{in \; }
{ \pgf (l_1 \otimes
\vec{k_1} \otimes \vec{k_2} \otimes l_2 \otimes q)}
\end{multline}
 where $\vec{k_i}$ denotes a tensor of $p$ elements of type $\kappa$:
$$\vec{k_i}=k_{i1}\otimes \dots \otimes k_{ip},$$
and
\begin{align*}
\mathtt{step} = 
\lambda k'' \lambda b\otimes \vec{k_1}\otimes l. &
            \mylet{(\mathtt{dup})k_{11}}{c\otimes d}{}\\
& \mathtt{true}\otimes k'' \otimes c \otimes k_{12}\dots \otimes k_{1(p-1)} \otimes \\
&  (\mbox{\tt{if} } b \mbox{ \tt{then} }        (g)d \mbox{ \tt{else} } I ) l
\end{align*}
 is of type $\mathtt{step} : !(\kappa \multimap
(Bool\otimes\kappa^p\otimes\beta \multimap
Bool\otimes\kappa^p\otimes\beta))$. The terms $\mathtt{base^+} = \pgf
(\mathtt{false}\otimes \star^p \otimes x')$ and $\mathtt{base^-} =
\pgf (\mathtt{false}\otimes \star^p \otimes x)$ are of type
$\mathtt{base^+} : \pgf (Bool\otimes\kappa^p\otimes\beta)$.


 Note that duplication of values of type $\kappa$ with $\mathtt{dup}$
has been used in the term $\mathtt{c2cw}$ (via $\mathtt{step}$) to build the window.
This only comes from the fact that a value written on the tape used
for an operation or a test remains written and can be used another time. 
This is actually the only place in the encoding where $\mathtt{dup}$ is used.

Now, once given a configuration with window, to perform a transition
step we need a term which will, depending on the state, select the
right action to perform:
$$\mathtt{next\_conf} : Q \multimap \alpha \otimes \kappa^p \otimes\kappa^p \otimes\alpha \multimap \alpha \otimes \alpha \otimes Q$$
This term simply uses the definition of states:
$$ \mathtt{next\_conf}= \la q. (q) t_1 \dots t_d,$$
where $t_j$ is a term corresponding to the action $\mu (q_j)$ of the transition table.
 To define the $t_j$s we have to consider the
three possible transitions in the BSS-machine:
\begin{enumerate}
\item (computation) the top of the positive part of the tape is replaced 
with the application of an operation $op_i$ to the first $p$ elements 
of the tape, $q'$ is the new state of the machine:
\begin{align*}
t_j =&
 \lambda l_1 \otimes \vec{k_1} \otimes \vec{k_2} \otimes l_2.\\  
 &((g)k_{11})l_1 \otimes \\
 & \otimes ((g)(\mathtt{op}_i)k_{21} \dots k_{2p})l_2 \otimes q'
\end{align*}

\item (branch) the branch case, the machine chooses the next state 
$q_1$ or $q_2$ depending on the result of the evaluation of the
 relation $\rho_i$ with the first $p$
elements of the positive tape as arguments
\begin{align*}
t_j =&
 \lambda l_1 \otimes \vec{k_1} \otimes \vec{k_2} \otimes l_2.  \\
 & ((g)k_{11})l_1 \otimes \\
 & \otimes ((g)k_{21})l_2 \otimes \\
 & \otimes \mbox{{\tt if} 
$(\mathtt{rho}_i)k_{21} \dots k_{2p}$ \tt{then} $q_1$ \tt{else} $q_2$}
\end{align*}
\item (shift) the left shift consists in moving the 
first element of the negative tape to the top of the positive
one with $q'$ as the new state:
\begin{align*}
t_j = &
\lambda l_1 \otimes \vec{k_1} \otimes \vec{k_2} \otimes l_2.  \\
 & l_1 \otimes\\
& \otimes ((g)k_{11}) ((g)k_{21}) l_2 \otimes q'
\end{align*}
analogously we do for the right shift:
\begin{align*}
t_j = &
\lambda l_1 \otimes \vec{k_1} \otimes \vec{k_2} \otimes l_2.  \\
 & ((g)k_{21}) ((g)k_{11})l_1 \otimes\\
& \otimes l_2 \otimes q'
\end{align*}
\end{enumerate}

 Finally, using the terms $\mathtt{c2cw}$ and $\mathtt{next\_conf}$ we define the term 
$\mathtt{c2c}$ which performs a transition step on a configuration:
\begin{multline} 
\mathtt{c2c} = \lambda c \lambda g \lambda x \lambda x'
\mathtt{let}
(\mathtt{c2cw})c g x x' \\ 
\mathtt{be \; }
\pgf ( l_1 \otimes \vec{k_1} \otimes \vec{k_2} \otimes l_2 \otimes q)\\
\mathtt{in \; }
{ \pgf( (\mathtt{next\_conf})q) (l_1 \otimes
\vec{k_1} \otimes \vec{k_2} \otimes l_2)}
\end{multline}

Note that $g,x,x'$ are bound by the $\mathtt{c2c}$ 
abstractions.

\subsection{Completing the encoding.}

 Just as in \cite{AspertiRoversi2002} one can define the following terms:
$$\begin{array}{ll}
\mathtt{length}: & \liste{\kappa} \fm \pgf N, \\
\mathtt{init}: & \liste{\kappa} \fm C, \\
\mathtt{extract}: & C \fm \liste{\kappa}.
\end{array}$$

The term $\mathtt{length}$ computes the length of the list as a tally
integer; $\mathtt{init}$ maps a list $l$ onto the corresponding
initial configuration $\langle nil, l, q_0\rangle$; $\mathtt{extract}$ recovers
from a configuration $\langle f^-, f^+, q\rangle$ the list corresponding to $f^+$.

Now, given an input $l$ of type $\liste{\kappa}$ for the machine, we
will need to use $l$ for two purposes:
\begin{enumerate}[(i)]
\item to produce the initial
configuration (with $\mathtt{init}$),
\item to yield an integer (its length) $n$, from which the time bound
for the machine will be computed.
\end{enumerate}

For that it is easy to define using $\mathtt{fold}$ a term
$\mathtt{Ilength}:\liste{\kappa} \fm \pgf(\liste{\kappa}\otimes N)$ such that:
$$(\mathtt{Ilength})l
\xreduce{\star} \pgf (l \otimes n),$$ 
where $n$ is the length of $l$.

Recall that:
\begin{lemma}[\cite{AspertiRoversi2002}]
 For any polynomial $P$ in $\NN [X]$ there exists an integer $k$
and a term $t_P: N \fm \pgf^k N$ such that $t_P$ represents $P$.
\end{lemma}
 Now, let $\mu$ be a polytime BSS machine with polynomial $P$.
One can define a term $u$ simulating  $\mu$ in the following way:
\begin{itemize}
\item apply $\mathtt{Ilength}$, and then $\mathtt{init}$ to the l.h.s.
  result to get a configuration $c_0$, and $t_P$ to the r.h.s. result
  to get an integer $m=P(n)$ (where $n$ is the length of the input);
\item use $m$ to iterate the term $\mathtt{c2c}$, $m$ times starting from $c_0$
  and get a configuration $c_1$;
\item apply $\mathtt{extract}$ to $c_1$.
\end{itemize}

Typing in a suitable way this procedure one obtains a term $u:
\liste{\kappa} \fm \pgf^d \liste{\kappa}$. We thus have:
\begin{theorem}
  For any function $f$ in $FP(\KK)$, there exists an integer $d$ and a
  term $u$ of $\lambdak$ with type $\liste{\kappa} \fm \pgf^d
  \liste{\kappa}$ representing $f$.
\end{theorem}

\section{Conclusions} 

We have presented an extension of light affine lambda-calculus to
computation on an arbitrary ring structure $\KK$. The definition of
this extension is quite natural and it characterizes the BSS class
$FP(\KK)$ in the same way light affine lambda-calculus characterized
the classical class $FP$. Compared to the characterization by safe
recursion from \cite{Marion-BSS} our approach offers the advantage of
integrating higher-order constructs which are likely to be useful in
describing numerical analysis algorithms. We plan to examine some
programming examples of algorithms in our calculus. It would also be
interesting to see if other calculi which characterize $FP$ in the
classical setting and have higher-order such as those of
\cite{Lafont02,Hofmann00,BellantoniNigglSchwichtenberg00} can be
extended in the same way to the BSS setting.

\bibliographystyle{alpha}
\bibliography{refBSS}
\end{document}